\documentstyle[preprint,prl,aps]{revtex}
\begin{document}
\tightenlines
\draft
\title{Electrostatics of Inhomogeneous Quantum Hall Liquid}
\author{A. L. Efros} 
\address{\vspace{-0.1cm} University of Utah, Salt Lake City, UT 84112 }
\maketitle
\begin{abstract}
The distribution of electron density  in the quantum Hall liquid 
 is considered in the presence of
macroscopic density gradient caused by side electrodes or inhomogeneous doping. In this case 
different Landau levels are occupied in different regions of a sample. These regions are separated by ``incompressible liquid''.
It is shown that  the applicability of the approach by Chklovskii {\it et al.} is substantially
restricted if the density gradient is not very large  and disorder is important. Due to the fluctuations of the remote donor's density
the liquid in the transition region can not be considered as completely incompressible. In the typical
situation, when the gap between Landau levels is not much larger than the energy of
 disorder, the transition region is 
 a wide  band  where electron density, averaged over
 the fluctuations, is independent
of magnetic field. The band   is a random mixture of regions occupied by electrons of upper level,
by holes of lower level and by incompressible liquid. The width of
 this band   is calculated and an
analytical expression for the fraction of incompressible liquid in different parts of this
band is given.  
\end{abstract}

\pacs{73.20.Dx,73.40.Hm}

\section{Introduction}
During the last few years the new methods have been developed  for  direct imaging  of the
charge and current of the quantum Hall liquid\cite{sha,23,24,ash,26,27,amir}. They are especially
 effective if the distribution changes along the sample either because of the sweeping magnetic
field or due to an  intentionally created
   gradient of electron density.
These  methods can be useful for studying a spatial modulation of the 
 charge density
 in the electron liquid, which appears due to
  electron-electron interaction\cite{koul,chal,rez,phi}.  Another very robust
and very simple reason for inhomogeneity of electron density  is fluctuations of charged donors
 in the
doped region of semiconductor. These fluctuations cause the changes of electron density which are
strongly dependent on the screening. On the other hand the screening decreases drastically each
time when the chemical potential passes through an integer or a fractional gap. Thus, both the
screening and the electron  density itself strongly depend on the magnetic field. To study
comparatively weak effects of  the ``internal'' modulation of the  liquid  one should first
have a possibility to discriminate the ``external'' effects. Moreover, the screening properties of the
quantum Hall liquid is an interesting problem  and imaging of the fluctuations induced by
donors may 
 provide the  new opportunities for quantitative understanding of this problem.

The theoretical study of the screening properties have been conducted by many
people\cite{ef1,bee,cha,chkl,ef4,jim1,pik}. In the first part of this paper I discuss 
 the hierarchy of lengths and magnitudes of the density fluctuations in the
macroscopically homogeneous
 liquid at different values of parameters.  These results were partially   obtained  in the
papers
\cite{ef4,pik}, however, since the local density fluctuations were not an experimental object
 at the
time, these
 fluctuations  were not
 properly described. I concentrate here on the thermodynamic density of states $dn/d\mu$.
 In fact,  fluctuations of the density make liquid compressible at all values of average 
density. This means that 
$dn/d\mu$ is non-zero even when the chemical potential
$\mu$ is inside the gap. Note  that from the  microscopic point of view the system with a
 long-range disorder is still a
 mixture of metallic and incompressible phases. Finite compressibility appears in the
 electrostatics, where electric field is averaged over the size larger than the sizes
 of typical fluctuations. Depending on  the sizes of the tips which are  used for imaging and the sizes
of fluctuations one can measure either average values, like $dn/d\mu$, or fluctuations themselves.

In the second part of the paper I consider a system with a
density gradient. In this case the average occupations of different Landau levels (LL's)
 depend on coordinate.
 The problem is that between the regions occupied by electrons of the two different LL's there is 
a region of ``incompressible liquid'', where density is supposed to be constant.
This problem has been previously addressed by   Chklovskii, Shklovskii and Glazman\cite{chkl}, referred
below as CSG. 
They  have presented 
an analytical solution of the equations of the  non-linear screening in a magnetic field
 completely ignoring disorder.   Their solution contains  a strip of incompressible liquid which
separates compressible regions of two different LL's. From electrostatic point of view this 
strip has  a dipole moment per length which depends on  magnetic field.

 I consider the case when  applied gradient of density  is much  smaller than the random
 gradients created by disorder. In this case transition region is   a random  mixture of different
 compressible phases  and incompressible phase. The system 
 is neutral in average if one takes into account a realistic
 value of $dn/d\mu$. However, there are strong fluctuations of density,  which 
 are 
  the only manifestation of the  
 transition  region.

\section{ Density fluctuations in macroscopically homogeneous system}
I  consider a model which consists of the plane with the two-dimensional electron liquid
 (TDEL) in a 
perpendicular magnetic field. Only few lowest LL's are supposed to be occupied.  The
parallel  plane with randomly distributed charged donors is at a distance $s$
 from the TDEL. The
average density of the charged donors is
$C$.  We assume  that $s$ is much larger than both the average distances between electrons and
the distance  between 
donors. All spatial
harmonics of the donor fluctuations with the wave length $R$ smaller than $s$ create
 an exponentially small
potential in the plane of the TDEL. The mean square  fluctuation of the donor charge in the square $s\times s$ is 
 $e\sqrt{Cs^2}$, while the density fluctuation is of order $\sqrt{C} /s$. Such fluctuations 
create a random  potential
of order of $W=e^2\sqrt{C}/\kappa$, where $\kappa$ is the lattice dielectric constant.
 This potential can be successfully screened by  by a small redistribution
 of electron  density if average  density $n$ of the highest partially occupied LL
is larger than $\sqrt{C} /s$.
This type of screening with $\delta n/n \ll 1$ is called the  linear screening.
 Because of the linear screening the
potential of the TDEL can be considered as a constant
 From electrostatic point
of view  the TDEL in this case is a metal, however,  $dn/d\mu$ is negative\cite{ef1,jime}.

 The ratio of the gap $\Delta$ to the energy $W$ is an   important parameter of the problem.
For integer gaps in GaAs-based
 structures $\Delta=W$ in magnetic field $B\approx 2{\rm T}$ 
 if  $C=10^{11}{\rm cm}^{-2}$. In what
follows it is convenient to consider separately  two different cases.

\subsection{ $\Delta \gg W$.}
In this subsection I refer to the work\cite{ef4}  where the non-linear screening has been considered in the
one-level approximation ignoring holes at the lower LL.  The  break down of the
linear screening has been found at
$n_c=0.42\sqrt{C} /s$.  It  happens very sharp
and at  lower density the chemical potential goes down as 
\begin{equation}
\mu=-{W\over 14} \left({C^{1/2}\over n s}-2.4\right)=-{WC^{1/2}\over 14s}\left({1\over n}-
{1\over n_c}\right),
\label{mu}
\end{equation}
where $n<n_c$.

 The  computer modeling  does not show any deviation from Eq.~(\ref{mu}) in the region defined by inequalities
$0>\mu>-2W$ and  $2.4>C^{1/2}/ n s >40$. 
Thus, if $\Delta<4W$, one can use Eq.~(\ref{mu}) to find the thermodynamic density of states $dn/d\mu$. 
Note that $d\mu/dn$ as obtained from Eq. (\ref{mu}) is positive. This approximation ignores
 internal energy of the TDEL which comes from the interaction and which is responsible for the 
 negative compressibility.
The chemical potential of an ideal interacting system can be just added to the right hand part of
Eq. (\ref{mu}). Then one can find that 
 the compressibility changes  sign
 with decreasing $n$ at $n\approx n_c$.

In the region of the non-linear screening the characteristic size $R$ of the potential
fluctuations should increase with decreasing density as $\sqrt{c}/n$. This result has been 
first obtained by Shklovskii and Efros\cite{book} for the  3d non-linear screening. It can be  easily
generalized  for the  2d-case as well. At $n=n_c$,
the size $R$ is of the order of linear screening radius. This length is supposed to be much smaller
 than all characteristic lengths of the non-linear screening, which start with the
 spacer length $s$. Assuming that the linear screening radius is    zero,
 one can propose the  equation
\begin{equation}
R={b\sqrt{c}\over n}(0.42-ns/\sqrt{c})={sb\over n}(n_c-n),
\label{Re}
\end{equation}
where $b$ is a numerical constant and $n<n_c$. For a very rough estimate of the  constant $b$ one
can analyze the size effect in computer simulation of the chemical potential\cite{ef4}
 at small $n$. Such an
analysis shows that $b$ is not very far from unity.

 At $\mu<-\Delta/2$ the screening by the
holes in the lower LL becomes important. One can rewrite Eq.~(\ref{mu}) in such a form  that it will be
applicable when $\mu$ is near the lower LL 
\begin{equation}
\mu=-\Delta+{W\over 14} \left({C^{1/2}\over p s}-2.4\right)=-\Delta +{WC^{1/2}\over 14s}\left({1\over p}-
{1\over n_c}\right),
\label{mup}
\end{equation}
where the density of holes  $p$ in the lower LL is less than $n_c$.

The Eqs.~(\ref{mu},\ref{mup}) are not exact in the middle of the gap. 
 Nevertheless, one can use them to  estimate the electron and
hole densities at $\mu=-\Delta/2$ as
\begin{equation}
n_m=p_m={e^2 C\over 7\kappa\Delta}.
\label{n}
\end{equation}

Electrostatic potential in the plane of TDEL changes with decreasing electron density $n$ in the 
following way.
 At $n>n_c$ a strong linear screening makes the change of the 
 potential negligible.
The peaks in the   potential appear when
$n<n_c$. In the planar regions
the energy equals $\mu$ and electron density is not constant. We call these regions metallic.
 In the region of incompressible liquid  the potential energy exceeds $\mu$.
The percolation through  metallic regions disappears at $n_p=0.11 \sqrt{C}/s$. At 
 $\mu=-\Delta/2$  almost all
the area is occupied by the incompressible liquid.  When $\mu$ becomes closer to the lower
 LL metallic regions
appear again but this time they consist of the holes in the lower  LL. The percolation through
 this new metal
appears at hole density $p_p=n_p$ and, at $p_c=n_c$,
 the linear screening transforms all the area into metal.

 In my early papers\cite{ef1}  I
predicted that the width of the plateaus of the quantum Hall effect in a clean material  should be  $2 n_p$.  
Later on it becomes clear that the low-temperature  experimental values  are much larger and
 this contradiction  is still not resolved completely. Koulakov {\it et al.}\cite{koul}
 argued that the plateaus become larger because of the pinning of modulated TDEL.
Hopefully, the simultaneous measurements of electron density distribution and quantum Hall effect may answer
this question.

\subsection{ $\Delta \ll W$.}
The distribution of electron density  in this case can be obtained in a framework of an
 analytical theory which has
been constructed  for the fractional gaps\cite{pik} to explain capacitance data by
 Eisenstein {\it et
al.}\cite{jime}. This theory can be reformulated for the integer gaps in the following way.

The non-linear screening appears at the same value of electron density in the upper LL, namely at
 $n_c=0.42\sqrt{C} /s$. With
decreasing $n$ the random potential becomes very soon of the order of $\Delta$.
 The plane with the TDEL
consists of  metallic regions of two different types.
  One type contains electron of the upper LL and the other one
contains holes of the lower LL. The typical size of this regions is of the order of
 the spacer width $s$. A simple
electrostatic estimate shows\cite{ef1,chkl} that the width of the incompressible strips
 $l$ between these two metals is
of the order of $s\sqrt{(\Delta/W)}$.  Since $l\ll s$, one can use the quantitative theory by CSG to
write
\begin{equation}
l^2={4\kappa \Delta\over {\pi^2 e^2}|\nabla n({\bf r})|} ,
\label{shk}
\end{equation}
where density gradient is taken at a boundary of a metallic region in the direction perpendicular to the boundary.
Since statistical properties of the donor distribution $C({\bf r})$  is known, one can calculate the distribution of
electron density. The most important result is the fraction $Q$ of the plane occupied by the
incompressible liquid. It is given by equation
\begin{equation}
Q=2(3/\pi^5)^{1/4}\Gamma ({5\over 4})\left({\Delta\over W}\right)^{1/2}\exp-{(n-\nu n_0)^2\over \delta^2},
\label{fra}
\end{equation}
where $\delta^2=C/(4 \pi s^2)$, and $\nu$ is an integer filling factor. The  incompressible
liquid occupies the maximum fraction of the plane  when average electron density $n$ is $\nu n_0$. In this case
$Q\approx 0.57(
\Delta/W)^{1/2}$. Computer simulation shows\cite{pik} that this result is valid at least at $ \Delta/W\leq 0.4$.
Note that at $\Delta/W= 0.4$ the maximum fraction of incompressible liquid $Q\approx 0.36$ 

Finally, at $\Delta<W$ the fraction of incompressible liquid is small even when chemical potential $\mu$ is
exactly in the middle of the gap. At this point the most part of the plane is occupied by electron and hole metals
with narrow filaments of incompressible liquid between them. The thermodynamic density of states 
 is of the order
$\Delta/n_c$.
\section{Spatial transition between Landau levels due to a  macroscopic density gradient.}

Suppose now that there is a small density gradient $g$ in $x$-direction created by the side
electrodes or  inhomogeneous distribution of donors.  We assume that  that this gradient is
much smaller than the typical  microscopic gradient of density $g_f$  that appears due to fluctuations of
donor's charge in the scale $s$. One gets
\begin{equation}
g_f=\sqrt{C}/s^2.
\end{equation}
The value of $g_f$ is $1.3\times 10^{16}{\rm cm}^{-3}$ at $C=10^{11}{\rm cm}^{-2}$ and $s=5\times
10^{-6}{\rm cm}$.

 If the thermodynamic density of states  is
large enough and
$g$ is small enough, one can consider the TDEL as a metal  where macroscopic electric
field is averaged over the distances larger than the size of the fluctuations described in the
previous section. The electron density can be found from the condition that electrostatic potential
$\phi$ is constant  in the plane of the
 TDEL
while the density itself is determined by a normal component of electric field $E_z$. In this
approximation density is independent of the the magnetic field and reproduces the density gradient of
charged donors. Since the TDEL is not an ideal metal this approximation is not exact. The
 lateral electric
field $E_x$ is non-zero  and it can be found from  
  the  condition  $\mu +e\phi=
{\rm const}$. Thus,  $E_x=
(1/e)(d\mu/dn)g$. The metallic approximation  for electron density  is good if
 $|E_x|\ll |E_z|$,
where $E_z=4\pi e n(x)/\kappa$ is a normal component of electric field and the density $ n(x)$ does
not include completely occupied LL's. The general condition of the metallic approximation is 
\begin{equation}
|{d\mu\over dn}|{g \kappa \over 4\pi e^2 n(x)}\ll 1.
\label{cond}
\end{equation}
This approximation is good at small gradients $g$ and at large $|dn/d\mu|$. The theory by CSG
 assumes $dn/d\mu=0$. Thus, the two theories are applicable at different conditions.

In the  case $\Delta<W$ one can  
substitute the estimate $d\mu/dn\approx \Delta/n_c$ into Eq.(\ref{cond}) to find that metallic 
approximation is valid if 
\begin{equation}
{g\over 4\pi g_f}{\Delta\over W}\ll 1.
\end{equation}
At $\Delta\le W$ this condition is always fulfilled. Note that CSG also mentioned that there theory is not applicable at  $\Delta< W$.

Now I come to the case of a strong magnetic field and weak disorder where  $\Delta\gg W$. 
 The electron density becomes inhomogeneous at $n<n_c$. At this density the estimate for the
 thermodynamic density of states 
is $dn/d\mu\approx n_c/W$, and  the metallic approximation works if $g\ll g_f$. This
 condition is usually fulfilled, but at lower density, when chemical potential is deep in the gap,
 the condition changes. The 
spatial point, where the chemical potential is in the middle of the gap, is the most dangerous for the
metallic approximation because of the small thermodynamic density of states. Thus, the metallic
approximation is valid everywhere if
Eq. (\ref{cond}) is fulfilled at the point
$\mu=-\Delta/2$, where  $n=n_m$. Making use of Eqs. (\ref{mu},\ref{n},\ref{cond} ) one gets the condition
\begin{equation}
2\left({\Delta \over W}\right)^3{g\over g_f}\ll 1.
\label{main}
\end{equation}
If the condition Eq. (\ref{main}) is not fulfilled
 the metallic
approximation is  violated within  some spatial region with the chemical potential
$|\mu-\Delta/2|<\Delta^\prime$.  In this case one should apply the theory by CSG 
 to the central part of the transition region substituting effective gap
 $\Delta^\prime$ instead of $\Delta$. With increasing gradient 
and 
increasing magnetic field the length of  this central part increases. Note that this possibility
has been considered by CSG in Fig. 5 of their paper.
\section{Conclusion}
 The theory by CSG gives a sharp transition strip  between two metallic regions corresponding two
LL's. Electron densities changes due to this transition to form the ``dipolar strip" with
 the dipole
moment depending on magnetic field. The width of the strip is given by Eq. (\ref{shk}) where 
$|\nabla n({\bf r})|$ should be substituted by $g$.

 I have shown above  that at small $g$ and moderate magnetic fields  the average
 electron density is the same as without
magnetic field. The fluctuations of this  density are the only manifestation of the transition.
These fluctuation are strong in the spatial region with the  non-linear screening. The size
 of this
region
$l$ can be obtained from condition  $g l=2 n_c$. This might be a large band  of the micron size.
 It is much larger than the typical size of the fluctuations. That is why I can use Eq.
 (\ref{cond}) with compressibility $d\mu/dn$ resulting from these fluctuations.
 Both the shape and the size of the fluctuations
strongly depend on magnetic field.

If $\Delta\le W$, the fraction of incompressible liquid as a function of coordinate  can be
 obtained  from the Eq.(\ref{fra}) by substituting $n=n(x)$, where $n(x)$ is electron
density without magnetic field.

One can compare this conclusion with the experimental data by Yakoby {\it et al.}\cite{amir},
where $\Delta<W$.
This data show wide transition regions with intermediate compressibility, which is probably, the 
result of averaging.
\section{Acknowledgment} 
I am grateful to A. Yakoby for attracting my attention to this problem 
 and for sending me the preprint of the
 paper\cite{amir}.

\end{document}